\documentclass{iau}
\usepackage{graphicx}
\usepackage[authoryear]{natbib}

\newcommand{\mdot}{~\ensuremath{\dot{m}}}
\newcommand{\Msun}{~\ensuremath{M_\odot}}
\newcommand{\Gpc}{~\ensuremath{\rm Gpc}}
\newcommand{\ergl}{~\ensuremath{\rm erg \, s^{-1}}}

\newcommand{\aap}{A\&A}
\newcommand{\apj}{Ap.~J}

\title[Quasar microlensing] 
{Microlensing evidence for super-Eddington disk accretion in quasars}

\author[Shakura \& Abolmasov]   
{ Nikolai I. Shakura
 \and  Pavel Abolmasov}
\affiliation{$^1$ Sternberg Astronomical Institute \\
 Universitetsky pr., 13, Moscow 119991, Russia \\ 
 email: {\tt pavel.abolmasov@gmail.com}}
\pubyear{2012}
\volume{290}  
\jname{Feeding compact objects: Accretion on all scales}
\editors{C.M. Zhang, T. Belloni, M. M\'endez \& S.N. Zhang, eds.}

\begin{document}

\maketitle

\begin{abstract}
Gravitational microlensing by the stellar population of lensing galaxies
provides an important opportunity to spatially resolve the accretion disk
structure in strongly lensed quasars. Some of the objects (like Einstein's
cross) are reasonably consistent with the predictions of the standard
accretion disk model. In other cases, the size of the emitting region is
larger than predicted by the standard thin disk theory and
practically independent on wavelength. This may be interpreted as an
observational manifestation of an
optically-thick scattering envelope possibly related to super-Eddington
accretion with outflows.
\end{abstract}

\firstsection 
\section{Introduction}

The optical/UV angular sizes of quasar accretion disks at $z\gtrsim 1 $ are $\sim
0.1\,\rm \mu as$.
%
%
These sizes may be resolved only by indirect methods such as microlensing by
the stellar population of lensing galaxies
\citep{wambsganss}. Fortunately, the Einstein-Chwolson radii of Solar-mass
microlenses at $\sim\Gpc$ distances are of the same order, and all the
microlensing curves are sensitive to the angular size of the source. 

Disk size estimates based on microlensing effects show several features that
are difficult to interpret in the framework of standard 
thin accretion disk theory. The
effect seems to depend on the black hole mass.
While massive black holes ($M\gtrsim 10^9\Msun$) have nearly standard disks
  (but possibly larger in size, see \citet{morgan10},
smaller black hole masses conform less with the standard model. For
  the latter, ``disk'' size is practically independent on wavelength for
  smaller-mass quasars \citep{blackburne}.

Having similar luminosities $L \sim 10^{47}\ergl$, 
lower- and higher-mass bright lensed quasars
have different Eddington luminosity limits. Moderately
super-Eddington sources should form a scattering envelope that increases the
apparent size of the object and makes the size dependence on wavelength
weaker. In more detail we address this model in our recent paper \citet{paper2}.

\section{Quasi-spherical envelope model}

Consider a spherically symmetric wind formed by a supercritical disk.  
It is reasonable to assume that at each radius
$R$ the terminate 
velocity of the outflowing wind is virial: $v_w = \beta \sqrt{GM/R}$,
where $\beta\sim 1$. 
%
The wind is effectively launched from the spherisation radius 
$R_{sph} \simeq  \frac{3}{2}\frac{GM}{c^2} \mdot $.
%
Condition $ \tau=1$ for the radial optical thickness 
leads to the following photosphere radius estimate:

\begin{equation}\label{E:r1}
\frac{R_1 c^2}{GM} \simeq   \frac{f_w}{\sqrt{2}} \mdot^{3/2}
\end{equation}

This may be both larger and smaller than the accretion disk radial scale:
$
R_{d} \simeq 10.2 \left(\frac{\lambda_{em}}{0.25\mu}\right)^{4/3}
  \left(\frac{M}{10^9\Msun}\right)^{-1/3}\mdot^{1/3}  \times \frac{GM}{c^2} 
$. 
Note also that this photosphere is ``grey'' in the sense 
its size does not depend on the wavelength. 

\begin{figure}
\begin{center}
 \includegraphics[height=2.0in]{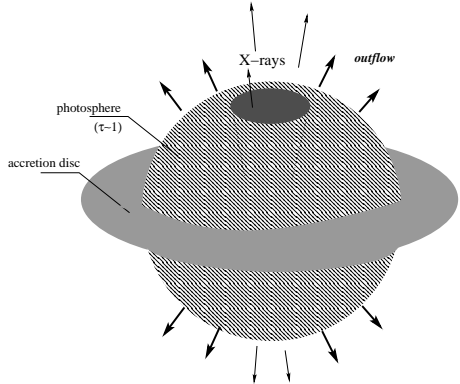}
 \caption{Qualitative sketch of the appearance of a super-Eddington quasar. }
   \label{fig:1}
\end{center}
\end{figure}

The issue of X-rays that are emitted by a considerably more compact region may
be addressed in a more comprehensive model taking into account deviations from
spherical symmetry. Both theory \citep{SS73} and simulations
\citep{ohsuga} predict that the wind formed by a supercritical accretion
disk should have a conical optically thin funnel. The X-ray emission
formed near the innermost stable orbit may reach the observer without
scattering or after a single reflection from the funnel wall. In both cases,
the size of the X-ray source will be considerably smaller than in the
optical/UV range. The appearance of the quasar photosphere is illustrated by
figure~\ref{fig:1}. 

\section{Summary}

We come to the conclusion that super-Eddington disk accretion may lead to
formation of a scattering pseudo-photosphere capable of explaining the spatial
properties of the emitting regions of distant lensed quasars in the optical/UV
range. Our work was supported by the RFBR grant 12-02-00186-a.



\begin{thebibliography}{}

\bibitem[\protect\citeauthoryear{{Abolmasov} \& {Shakura}}{{Abolmasov} \&
  {Shakura}}{2012}]{paper2}
{Abolmasov} P.,  {Shakura} N.~I.,  2012, ArXiv e-prints (arXiv:1208.1678)

\bibitem[\protect\citeauthoryear{{Blackburne}, {Pooley}, {Rappaport} \&
  {Schechter}}{{Blackburne} et~al.}{2011}]{blackburne}
{Blackburne} J.~A.,  {Pooley} D.,  {Rappaport} S.,    {Schechter} P.~L.,  2011,
  \apj, 729, 34

\bibitem[\protect\citeauthoryear{{Morgan}, {Kochanek}, {Morgan} \&
  {Falco}}{{Morgan} et~al.}{2010}]{morgan10}
{Morgan} C.~W.,  {Kochanek} C.~S.,  {Morgan} N.~D.,    {Falco} E.~E.,  2010,
  \apj, 712, 1129

\bibitem[\protect\citeauthoryear{{Ohsuga}, {Mori}, {Nakamoto} \&
  {Mineshige}}{{Ohsuga} et~al.}{2005}]{ohsuga}
{Ohsuga} K.,  {Mori} M.,  {Nakamoto} T.,    {Mineshige} S.,  2005, \apj, 628,
  368

\bibitem[\protect\citeauthoryear{{Shakura} \& {Sunyaev}}{{Shakura} \&
  {Sunyaev}}{1973}]{SS73}
{Shakura} N.~I.,  {Sunyaev} R.~A.,  1973, \aap, 24, 337

\bibitem[\protect\citeauthoryear{{Wambsganss}}{{Wambsganss}}{2006}]{wambsganss}
{Wambsganss} J.,  2006, in {G.~Meylan, P.~Jetzer, P.~North, P.~Schneider,
  C.~S.~Kochanek, \& J.~Wambsganss} ed., Saas-Fee Advanced Course 33:
  Gravitational Lensing: Strong, Weak and Micro {Part 4: Gravitational
  microlensing}.
pp 453--540

\end{thebibliography}

\end{document}